\documentclass[epj]{webofc}
\usepackage[utf8]{inputenc}
\usepackage[varg]{txfonts}   % Web of Conferences font
\usepackage{booktabs}
\usepackage{xcolor}
\usepackage{braket}
\definecolor{darkred}{rgb}{0.4,0.0,0.0}
\definecolor{darkgreen}{rgb}{0.0,0.4,0.0}
\definecolor{darkblue}{rgb}{0.0,0.0,0.4}
\usepackage[bookmarks,linktocpage,colorlinks,
    linkcolor = darkred,
    urlcolor  = darkblue,
    citecolor = darkgreen]{hyperref}
\usepackage{subfigure}
\wocname{EPJ Web of Conferences}
\woctitle{Lattice2017}
\begin{document}
%%%%%%%%%%%%%%%%%%%%%%%%%%%%%%%%%%%%%%%%%%%%%%%%%%%%%%%%%%%%%%%%%%%%%%%%%%%%%
%
\selectlanguage{english}
%----------------------------------------------------------------------------
\title{%
Satisfying positivity requirement in the Beyond Complex Langevin approach
}
%----------------------------------------------------------------------------
\author{%
\firstname{Adam} \lastname{Wyrzykowski}\inst{1}\fnsep\thanks{Speaker, \email{adam.wyrzykowski@doctoral.uj.edu.pl}} \fnsep\thanks{Acknowledges financial support by the NCN grant: UMO-2016/21/B/ST2/01492.} \and
\firstname{B\l a\.{z}ej} \lastname{Ruba}\inst{1} \fnsep\thanks{Acknowledges financial support by the NCN grant: UMO-2016/21/B/ST2/01492.}
% etc.
}
%----------------------------------------------------------------------------
\institute{%
M. Smoluchowski Institute of Physics, Jagiellonian University, Krakow, Poland
}
%----------------------------------------------------------------------------
\abstract{%
The problem of finding a positive distribution, which corresponds to a given complex density, is studied. By the requirement that the moments of the positive distribution and of the complex density are equal, one can reduce the problem to solving the matching conditions. These conditions are a set of quadratic equations, thus Groebner basis method was used to find its solutions when it is restricted to a few lowest-order moments. For a Gaussian complex density, these approximate solutions are compared with the exact solution, that is known in this special case. 

}
%----------------------------------------------------------------------------
\maketitle
%----------------------------------------------------------------------------
\section{Introduction}\label{sec-1}
Our goal is to represent integrals of complex densities over a real measure by integrals of real, positive probability distributions over a complex measure. Complex Langevin method \cite{parisi, klauder} has become a popular, and in many cases successful \cite{weingarten,salcedo,18,20}, approach to perform this task. However, in some situations problems with its convergence were observed \cite{6,8,sinclair}, so this topic attracted further investigation \cite{10,11,13,14,seilerproc} and also other methods were developed. The aim of this paper is analysis of a one-dimensional problem \cite{wosiek, seilerwosiek,wosiekproc,ruba} to find a positive and normalizable probability distribution $P(x,y)$ such that:
\begin{equation}
\int f(x)\rho(x)\ dx=\iint f(x+iy)P(x,y)\ dx dy
\label{1}
\end{equation}
for a given complex function $\rho(x)$ and every $f(x)$. This approach to the sign problem, although related to the Complex Langevin method \cite{parisi, klauder}, does not require introduction of the corresponding stochastic process. Instead, this method focuses only on satisfying the so-called matching conditions which follow from eq. (\ref{1}). Writing eq. (\ref{1}) for the moments, one obtains the following set of conditions:
\begin{equation}
\int x^r\rho(x)\ dx\equiv M_r=\iint (x+iy)^rP(x,y)\ dxdy\ .
\label{2}
\end{equation}
A proposal how this infinite set of integral equations can be solved is presented in Section \ref{sec-2}. In particular, a simple trick to satisfy positivity of $P(x,y)$ and numerical methods to solve resulting sets of polynomial equations are proposed. In Section \ref{sec-3}, the numerical results for Gaussian and quartic actions are presented. The summary is provided in Section \ref{sec-4}.  

\section{Matching conditions}\label{sec-2}
\subsection{Satisfying positivity}\label{sec-21}
In order to satisfy positivity of the probability distribution, one assumes that $P(x,y)=|\psi(x,y)|^2$. Then:
\begin{equation}
M_r=\iint\psi(x,y)^*(x+iy)^r\psi(x,y)\ dxdy=\braket{\psi|(x+iy)^r|\psi},
\label{3}
\end{equation}
where we employ the Dirac notation and think of $\psi(x,y)$ as a wavefunction of some quantum system in two dimensions. After expanding $\ket{\psi}$ in a basis:
\begin{equation}
\psi(x,y)=\sum_{m, n=0}^\infty c_{mn}\psi_m(x)\psi_n(y),
\label{4}
\end{equation}
e.g. the basis of two non-interacting harmonic oscillators, one obtains the following set of matching conditions:
\begin{equation}
M_r=\sum_{m'n'mn}c_{m'n'}c_{mn}\braket{\psi_{m'n'}|(x+iy)^r|\psi_{mn}},
\label{5}
\end{equation}
where $c_{mn}\in\mathbb{R}$ is assumed for simplicity. Using the properties of the harmonic oscillator basis, one can evaluate $\braket{\psi_{m'n'}|(x+iy)^r|\psi_{mn}}$, so these equations are second degree polynomials in coefficients $c_{mn}$. A few lowest-order equations are:
\begin{multline}
1=\sum_{mn}c_{mn}^2,\quad \text{Re }M_1=\sqrt{2}\sum_{mn}\sqrt{m+1}c_{m+1,n}c_{mn},\quad \text{Im }M_1=\sqrt{2}\sum_{mn}\sqrt{n+1}c_{m,n+1}c_{mn}\\
\text{Re }M_2=\sum_{mn}\left[(m-n)c_{mn}+\sqrt{(m+1)(m+2)}c_{m+2,n}-\sqrt{(n+1)(n+2)}c_{m,n+2}\right]c_{mn},\qquad \text{etc.}
\label{6}
\end{multline}
The quantity of interest in this approach is the probability distribution $|\psi(x,y)|^2$, not the "wavefunction" $\psi(x,t)$ itself. An interesting side remark is that it bears resemblance to the Nelson approach to quantum mechanics \cite{nelson}. This similarity is interesting from theoretical point of view and may be of profit for the numerical analysis of the problem, when searching for efficient tools to solve it.

\subsection{Gr\"{o}bner basis method}\label{sec-2extra}
Each of equations in the set (\ref{5}) includes infinitely many $c_{mn}$'s. Approximate solutions to such a set of equations can be found by introducing a cutoff on $(m,n)$. That is one leaves only a finite number of variables $c_{mn}$ (e.g. such that $m+n\leq N$) and an equal number of equations. Thus, the initial problem can be reduced to solving finite sets of second degree polynomial equations. The most general approach to such a problem is Gr\"{o}bner basis method (see e.g. \cite{groebner}), which provides an algorithm to find common roots of any set of polynomial equations. Finding a Gr\"{o}bner basis of a given set of equations 
%simplifies
can be used to simplify it to the form where the variables are succesively eliminated, i.e. such that variables $x_1,..., x_{i-1}$ do not appear in the $i$-th equation.\\
For a brief introduction to Gr\"{o}bner bases one needs to define a few concepts. First, a \emph{monomial ordering} must be specified, which is a non-unique task for multivariate polynomials. The most elementary and very useful choice is \emph{lexicographic} order:
\begin{equation}
x^\alpha>x^\beta\quad:\Leftrightarrow\quad\text{leftmost, nonzero entry in the vector }(\alpha-\beta)\text{ is positive},
\label{x1}
\end{equation}  
where $\alpha=(\alpha_1,\alpha_2,...,\alpha_n)$ is a vector composed of the exponents of the first monomial and $x^\alpha$ is a shortcut for $x_1^{\alpha_1}x_2^{\alpha_2}...x_n^{\alpha_n}$. 
%Leading term $\text{LT}(f)$ of a multivariate polynomial $f$ is defined as the term corresponding to the largest monomial (thus, it depends on the choice of ordering).
\emph{Multidegree} of a multivariate polynomial $f$ is defined as the vector $\alpha\in\mathbb{Z}_0^n$ corresponding to the largest monomial which appears in $f$ with a non-zero coefficient. The \emph{leading monomial} is $\text{LM}(f)=x^{\text{multideg}(f)}$, and the \emph{leading term} $\text{LT}(f)$ is the leading monomial together with its numerical coefficient. Clearly, all these quantities depend on the choice of ordering.\\
One also needs to define the division algorithm for multivariate polynomials \cite{groebner}:\\

\textbf{Theorem } Let us fix a monomial ordering. Given a multivariate polynomial $f\in K[x_1, ..., x_n]$ and a set of multivariate polynomials $F=\{f_1, ..., f_s\}\in K[x_1, ..., x_n]^s$, there exists a representation:
\begin{equation}
f=a_1f_1+a_2f_2+...+a_sf_s+r
\label{y1}
\end{equation}
of the following properties:
\begin{itemize}
\item $a_1, ..., a_s, r\in K[x_1, ..., x_n]$;
\item none of the monomials in $r$ is divisible by any of $\text{LT}(f_i)$;
\item $\text{multideg}(f)\geq\text{multideg}(a_if_i)$ for all non-zero $a_if_i$.
\end{itemize}

$r$ is called the remainder of $f$ on division by $F$. For example, since:
\begin{equation}
x^2y+xy^2+y^2=(x+y)(xy-1)+1\cdot(y^2-1)+x+y+1,
\label{y2}
\end{equation}
the remainder of $f=x^2y+xy^2+y^2$ on division by $F=\{xy-1, y^2-1\}$ is $r=x+y+1$.\\
The definition of a Gr\"{o}bner basis if following:\\

\textbf{Def. } Given a set of polynomials $F=(f_1, f_2, ..., f_s)\in K[x_1, ..., x_n]^s$, a set $G=(g_1, g_2,..., g_t)\in K[x_1, ..., x_n]^t$ is its Gr\"{o}bner basis if and only if all leading monomials of linear combinations of polynomials in $F$:
\begin{equation}
\sum_{i=1}^s h_if_i\qquad ;\ h_1, h_2,..., h_s\in K[x_1,...,x_n]
\label{x2}
\end{equation}
are divisible by at least one of $LM(g_i)$.\\

Notice, that in this context, a \emph{linear combination} is the sum of \emph{polynomial} multiples of the elements of $F$. The most common procedure to find a Gr\"{o}bner basis is the Buchberger's algorithm. An important auxillary quantity in this algorithm is $S$\emph{-polynomial}:
\begin{equation}
S(f,g)=\frac{x^\gamma}{LT(f)}\cdot f-\frac{x^\gamma}{LT(g)}\cdot g,
\label{x3}
\end{equation}
where $x^\gamma$ is the least common multiple of leading monomials in $f$ and $g$. S-polynomials are constructed so that the leading terms of the two polynomials cancel each other out.\\ In the first step of the Buchberger's algorithm, one takes the Gr\"{o}bner basis candidate to be equal to the initial set of polynomials $F=(f_1, ..., f_s)$:
\begin{equation}
G:=F.
\label{x4}
\end{equation}
In the next step, for every two polynomials $p$ and $q$ in $G$, the remainder of $S(p,q)$ on division by $G$ is calculated:
\begin{equation}
S:=\overline{S(p,q)}^G.
\label{x5}
\end{equation}
If this quantity is non-zero, i.e. if $S(p,q)$ is not divisible by the set $G$, $S$ is added to the Gr\"{o}bner basis candidate $G$. This procedure is repeated until $S(p,q)$ is divisible by $G$ for every $p, q\in G$. It can be summarized as follows \cite{groebner}:\\
\leavevmode{\parindent=1em\indent}\texttt{Input: }$F=(f_1, ..., f_s)$\\
\leavevmode{\parindent=0.8em\indent} \texttt{Output: a Gr\"{o}bner basis }$G=\{g_1, ...,g_t\}$, 
$F\subset G$\\
\newline
\leavevmode{\parindent=1em\indent}$G:=F$\\
\leavevmode{\parindent=1em\indent}\texttt{REPEAT}\\
\leavevmode{\parindent=2em\indent}$G':=G$\\
\leavevmode{\parindent=2em\indent}\texttt{FOR each pair }$\{p,q\}$, $p\neq q$ in $G'$\texttt{ DO}\\
\leavevmode{\parindent=3em\indent}$S:=\overline{S(p,q)}^{G'}$\\
\leavevmode{\parindent=3em\indent}\texttt{IF }$S\neq0$\texttt{ THEN }$G:=G\cup S$\\
\leavevmode{\parindent=1em\indent}\texttt{UNTIL }$G=G'$.\\

Example:\\

The Gr\"{o}bner basis method can be used to solve the set of equations:
\begin{equation}
\begin{cases}
f_1=x^2y+y=0\\
f_2=xy^2+x=0
\end{cases}.
\label{x6}
\end{equation}
Following the notation used in this section, $F=\{x^2y+y,xy^2+x\}$ and the initial Gr\"{o}bner basis cadidate is $G=\{g_1,g_2\}=\{x^2y+y,xy^2+x\}$. In the first step, one obtains $S(g_1,g_2)=x^2-y^2$, which is not further divisible by $G'$, so $\overline{S(g_1,g_2)}^{G'}=x^2-y^2$ and $G$ must be expanded by $g_3=x^2-y^2$. Similarly in the next step of the algorithm, $S(g_1,g_3)=\overline{S(g_1,g_3)}^{G'}=-y-y^3$ and $S(g_2,g_3)=-x^2-y^4=-g_3-y^2-y^4$, thus $\overline{S(g_2,g_3)}^{G'}=-y^2-y^4$. Hence, $G:=G\cup\{g_4,g_5\}=G\cup\{-y-y^3,-y^2-y^4\}$. One can check that after this extension $\overline{S(g_i,g_j)}^{G}=0$ for all elements of $G$. Therefore, the polynomials in the initial set of equations (\ref{x6}) can be replaced by the polynomials of the Gr\"{o}bner basis:
\begin{equation}
\begin{cases}
g_1=x^2y+y=0\\
g_2=xy^2+x=0\\
g_3=x^2-y^2=0\\
g_4=-y-y^3=0\\
g_5=-y^2-y^4=0
\end{cases}.
\label{x7}
\end{equation}
$g_1=yg_3-g_4$ and $g_5=yg_4$, so the first and the fifth equations are redundant, because they are satisfied due to the remaining ones. Also the Gr\"{o}bner basis can be reduced to $G=\{g_2,g_3,g_4\}$, since $g_1$ and $g_5$ are generated by $g_3$ and $g_4$. The fourth equation involves only $y$ and can be easily solved, $y\in\{0,i,-i\}$. After substituting these values to the second and third equations, one finally finds all solutions, $(x,y)\in\{(0,0),(i,i),(i,-i),(-i,i),(-i,-i)\}$.
\subsection{Solving matching conditions}\label{sec-22}
The facts, that Gr\"{o}bner basis method allows to find entire sets of solutions and it is exact, are its largest advantages. On the other hand, the main drawback is that the numerical complexity increases rapidly with the number of equations and variables \cite{dube}. We have therefore applied another, approximate procedure for larger sets of equations. The approximate algorithm used in this paper is a numerical search for the local minima of the sum:
\begin{equation}
\sum_i\left(LHS_i-RHS_i\right)^2
\label{7}
\end{equation}
with respect to the parameters $c_{00}, c_{01}, c_{10}, ..., c_{\text{cutoff}}$. $LHS_i$ and $RHS_i$ are the left and the right hand side of $i$-th equation of a given set of polynomial equations. The starting point of the minimization procedure is chosen randomly in $k$-dimensional space ($k$ is the number of variables). If the minimization procedure leads to a point $(c_{00}, c_{01}, c_{10}, ..., c_{\text{cutoff}})$ for which the goal function (\ref{7}) equals zero, this solution is also a solution to the set of polynomial equations. Then, this procedure is repeated for a large enough number of starting points to obtain all solutions of the problem.

\section{Results}\label{sec-3}
\subsection{Gaussian case}\label{sec-31}
\begin{figure}[thb]
   \centering
   	\includegraphics[width=10cm]{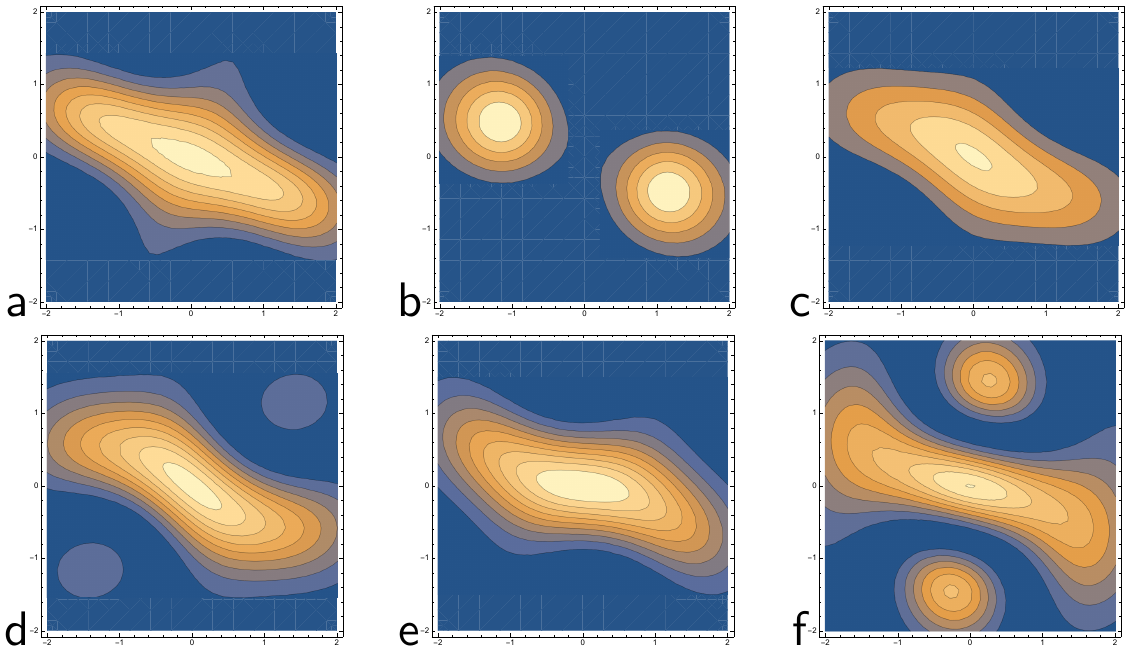}
	\caption{Contour plots of the probability distribution $P(x,y)$ corresponding to the different solutions of the system of equations (\ref{10}).
}
	\label{fig1}
\end{figure}

The two methods proposed in the previous section were used to find a positive probability distribution $P(x,y)$ in the Gaussian case:
\begin{equation}
\rho(x)=\sqrt{\frac{\sigma}{2\pi}}\text{e}^{-\sigma x^2/2}.
\label{8}
\end{equation}
For numerical calculations Wolfram Mathematica 10 is used (procedures \texttt{GroebnerBasis[]} and \texttt{FindMinimum[]}). Our results were compared with the exact solution, which in this case reads \cite{ambjornyang, wosiek}:
\begin{equation}
P_g(x,y)=\frac{\sigma_R\sqrt{1+r^2}}{\pi}\exp\left(-\sigma_R(x^2+2rxy+(1+2r^2)y^2)\right),
\label{9}
\end{equation}
where $\sigma_R=\text{Re }\sigma$, $\sigma_I=\text{Im }\sigma$ and $r=\sigma_R/\sigma_I$.\\
We introduce a cutoff $m+n\leq 4$, neglect $c_{mn}$ for odd $m+n$ and take the following symmetry assumptions: $c_{20}=-c_{02}$, $c_{40}=c_{04}$ and $c_{31}=-c_{13}$. Such a choice of symmeties is dictated by the behaviour of the coefficients $c_{mn}$ for the exact solution (\ref{9}). One obtains a system of 6 second degree equations for 6 variables:
\begin{equation}
\begin{cases}
1=c_{00}^2+2c_{02}^2+c_{11}^2+2c_{04}^2+2c_{13}^2+c_{22}^2\\
\frac{1}{\sqrt{2}}=-2\sqrt{2}c_{00}c_{02}-2\sqrt{6}c_{11}c_{13}-4\sqrt{3}c_{02}c_{04}+2\sqrt{2}c_{02}c_{22}\\
-\frac{1}{\sqrt{2}}=2c_{00}c_{11}+4c_{11}c_{22}+4\sqrt{3}c_{02}c_{13}\\
0=12c_{02}^2-6c_{11}^2+36c_{04}^2-6c_{00}c_{22}-18c_{22}^2+2\sqrt{6}c_{00}c_{04}-12\sqrt{6}c_{04}c_{22}\\
-3=-4\sqrt{6}c_{00}c_{13}-72c_{04}c_{13}-12\sqrt{2}c_{02}c_{11}-12\sqrt{6}c_{22}c_{13}\\
-\sqrt{2}=8\sqrt{6}c_{11}c_{13}-8\sqrt{3}c_{02}c_{04}+12\sqrt{2}c_{02}c_{22}
\end{cases},
\label{10}
\end{equation}
where, on the left hand sides, the moments $M_r=\int x^r\rho(x)\ dx$ are calculated for $\sigma=(1+i)/\sqrt{2}$. There are 12 solutions of this system of equations. The results provided by Gr\"{o}bner basis method are presented and compared to the exact solution (\ref{9}) in Tab. \ref{tab1}. Corresponding contour plots of $P(x,y)$ are depicted in Fig. \ref{fig1}.
\begin{table}[thb]
  \small
  \centering
  \caption{Solutions of the system of equations (\ref{10}) (remaining 6 solutions are minus these ones).}
   \begin{tabular}{lllllll}\toprule
&$c_{00}$&$c_{02}$&$c_{11}$&$c_{04}$&$c_{13}$&$c_{22}$\\ \midrule
a& \textbf{0.827}& \textbf{-0.154}& \textbf{-0.217}& -0.00815& 0.332& -0.00665\\
b& 0.520& -0.404& -0.572& -0.129& 0.126& -0.105\\
c& 0.855& -0.101& -0.382& 0.0608& 0.215& -0.0627\\
d& 0.738& -0.0817& -0.587& 0.145& 0.150& -0.104\\
e& -0.811& 0.339& 0.114& 0.00705& -0.190& -0.166\\
f& -0.431& 0.523& 0.376& -0.145& 0.0145& -0.290\\
from eq. (\ref{9}) & \textbf{0.910}&\textbf{-0.189}&\textbf{-0.267}&0.0478&0.0956&0.0390\\ \bottomrule
 \end{tabular}
\label{tab1}
\end{table}

\begin{figure}[thb]
   \centering
   	\includegraphics[width=13cm]{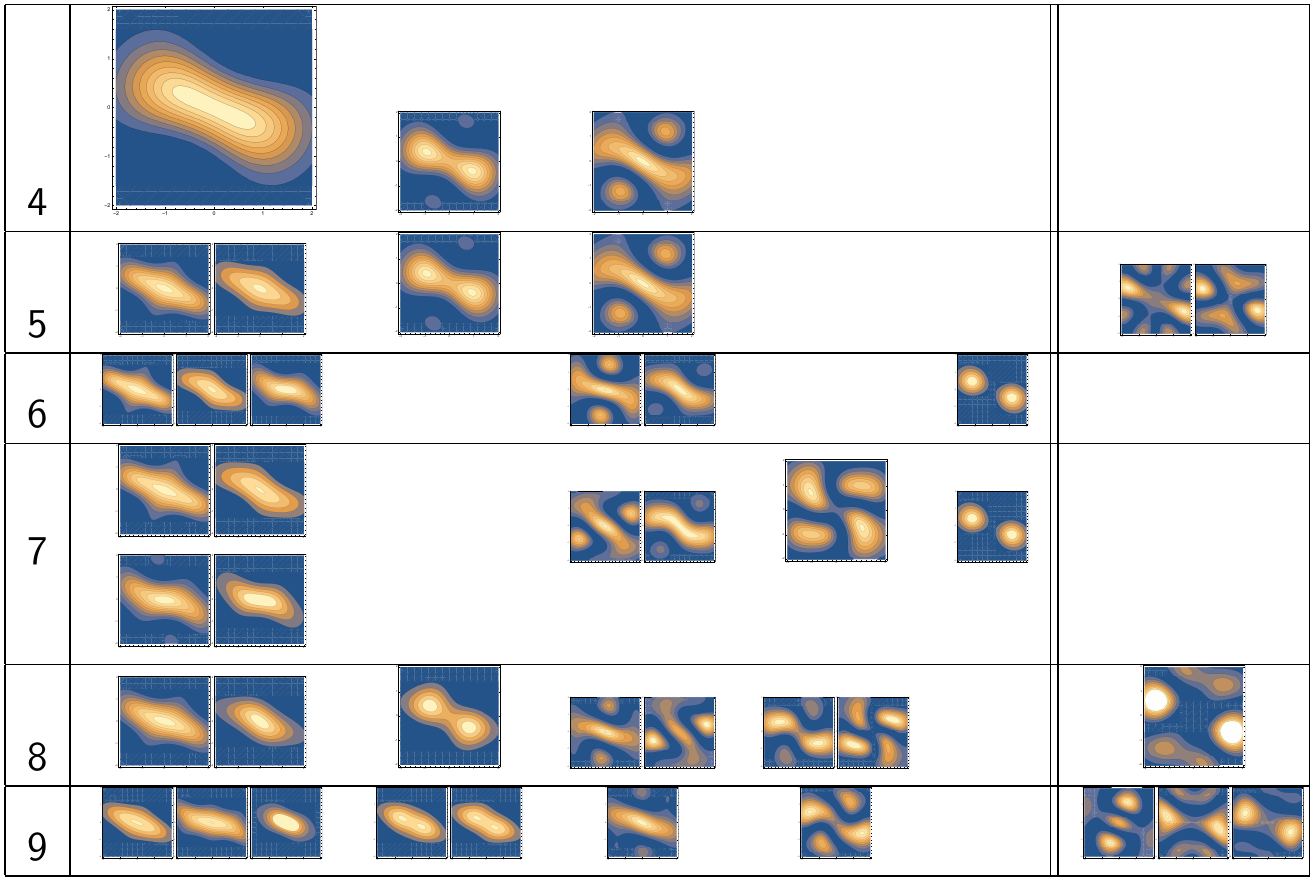}
	\caption{Dependence of solutions in the Gaussian case on the cutoff. The rows are numbered with the number of matching conditions and the solutions are arranged according to their respective proximity.
%Classification of solutions in Gaussian case with respect to the number of matching conditions.
}
\label{fig2}
\end{figure}
The most important observation from these calculations is high non-uniqueness of the problem. Both Gr\"{o}bner method and minimization method lead to a large number of solutions which increases with the number of matching conditions. Nevertheless, the solutions can be classified according to their stability with increasing the cutoff as shown in Fig. \ref{fig2}, where for every cutoff, the solutions are arranged into classes of solutions which are close to each other. A measure of distance used to determine proximity of solutions was:
\begin{equation}
||P_1-P_2||_2=\int_{-\infty}^\infty\int_{-\infty}^\infty \left(P_1(x,y)-P_2(x,y)\right)^2\ dxdy.
\label{a1}
\end{equation}
In particular, there exists a stable solution present in all steps of the algorithm and other, unstable solutions.

\subsection{Quartic case}\label{sec-32}

\begin{figure}[thb]
   \centering
   	\includegraphics[width=13cm]{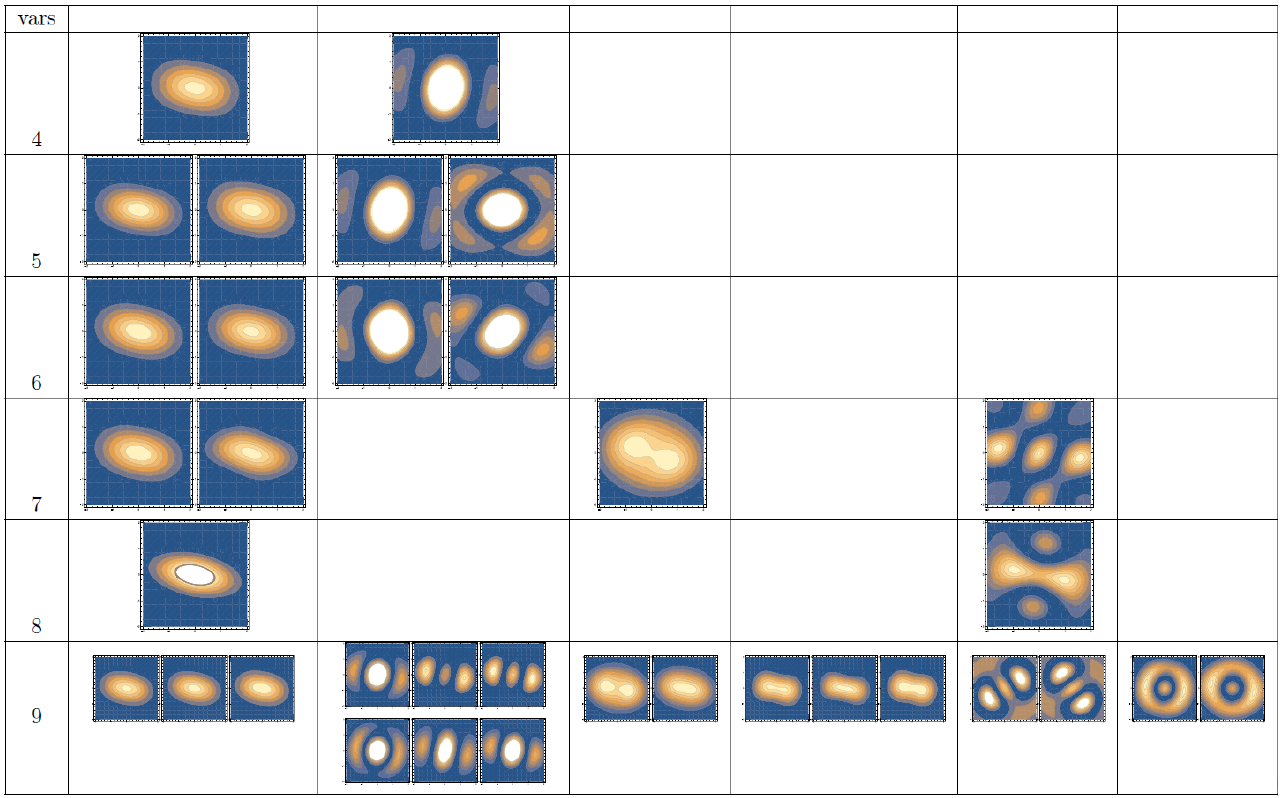}
	\caption{Classification of solutions for quartic action.}
\label{fig3}
\end{figure}

The same method can be applied to any other function $\rho(x)$. The only difference will be the values of the moments on the RHS of eq. (\ref{5}). Therefore, one can study the case of a quartic action:
\begin{equation}
\rho(x)=\frac{(8\lambda)^{1/4}}{\Gamma(1/4)}e^{-\lambda x^4/2}.
\label{11}
\end{equation}
For numerical calculations, we have taken $\lambda=(1+i)/\sqrt{2}$. Similarly to the previous case, Gr\"{o}bner basis method is efficient for no more than 6-8 variables, while for larger systems of equations the minimization method was used. The resulting positive distributions $P(x,y)$ and classification of those solutions are presented in Fig. \ref{fig3}. Again, a class of solutions which are present independently of the cutoff is observed (the first column in Fig. \ref{fig3}). Other solutions repeat for different cutoffs or appear only for a particular number of matching conditions, but are not totally cutoff-independent. A reasonable conjecture is that this class remains stable also for larger number of equations and these solutions approximate an exact solution to the full set of equations (\ref{5}).

\section{Summary}\label{sec-4}

In order to enforce positivity of the probability distribution it is assumed that $P(x,y)=|\psi(x,y)|^2$. This approach requires solving sets of second degree polynomial equations, which can be done with Gr\"{o}bner bases method. In this method, one obtains exact solutions of the matching conditions and with this approach all solutions are found. There are also other methods to approach this problem, for which satisfying positivity is not so simple, but which are linear and thus their mathematical treatment is much more successful (see e.g. \cite{ruba}).\\
Gr\"{o}bner bases method and the minimization method are used to find solutions for Gaussian and quartic actions. Non-uniqueness of the problem is directly observed in both cases. In Gaussian case, the known, exact solution is observed with the methods proposed in this paper. For both Gaussian and quartic actions, approximate solutions are found and classified according to their respective proximity. There are stable solutions which appear for every cutoff, as well as unstable ones. The next important challenge would be finding more efficient algorithms to obtain Gr\"{o}bner bases, since the numerical complexity increases rapidly with the number of equations.

%%%%%%%%%%%%%%%%%%%%%%%%%%%%%%%%%%%%%%%%%%%%%%%%%%%%%%%%%%%%%%%%%%%%%%%%%%%%%
\end{document}